\documentclass{article}

\newcommand{\angstrom}{\mbox{\normalfont\AA }}

\usepackage{arxiv}

\usepackage[utf8]{inputenc} 
\usepackage[T1]{fontenc}    
\usepackage{hyperref}       
\usepackage{url}            
\usepackage{booktabs}       
\usepackage{amsfonts}       
\usepackage{nicefrac}       
\usepackage{microtype}      
\usepackage{lipsum}

\title{Adsorption of \ce{H2} on Amorphous Solid Water Studied with Molecular Dynamics Simulations }

\usepackage[super,sort&compress,comma]{natbib} 
\usepackage[version=3]{mhchem}
\usepackage{mathptmx}
\usepackage{sectsty}
\usepackage{graphicx} 
\usepackage{lastpage}
\usepackage{gensymb}
\usepackage{adjustbox}
\usepackage[format=plain,justification=justified,singlelinecheck=false,font={stretch=1.125,small,sf},labelfont=bf,labelsep=space]{caption}
\usepackage{float}
\usepackage{fancyhdr}
\usepackage{fnpos}
\usepackage[english]{babel}
\addto{\captionsenglish}{%
  
}

\newcommand{\fig}{Fig.}

\newcommand{\figref}[1]{\fig~\ref{#1}}

\newcommand{\tabref}[1]{Table~\ref{#1}}

\usepackage{array}
\usepackage{charter}
\usepackage[T1]{fontenc}
\usepackage[usenames,dvipsnames]{xcolor}
\usepackage{setspace}
\usepackage[compact]{titlesec}
\usepackage{hyperref}

\author{
  Germ\'an Molpeceres \\
  Institute for Theoretical Chemistry\\
  University of Stuttgart\\
  \texttt{molpeceres@theochem.uni-stuttgart.de} \\
   \And
 Johannes K\"astner \\
Institute for Theoretical Chemistry\\
  University of Stuttgart\\
  \texttt{kaestner@theochem.uni-stuttgart.de} \\
}

\begin{document}
\maketitle

\begin{abstract}
We investigated the behavior of \ce{H2}, main constituent of the gas phase in dense clouds, after collision with amorphous solid water (ASW) surfaces, one of the most abundant chemical species of interstellar ices. We developed a general framework to study the adsorption dynamics of light species on interstellar ices. We provide binding energies and their distribution, sticking probabilities for incident energies between 1~meV and 60~meV, and thermal sticking coefficients between 10 and 300~K for surface temperatures from 10 to 110~K. We found that the sticking probability depends strongly on the adsorbate kinetic energy and the surface temperature, but hardly on the angle of incidence. We observed finite sticking probabilities above the thermal desorption temperature. Adsorption and thermal desorption should be considered as separate events with separate time scales. Laboratory results for these species have shown a gap in the trends attributed to the differently employed experimental techniques. Our results complement observations and extend them, increasing the range of gas temperatures under consideration. We plan to employ our method to study a variety of adsorbates, including radical and charged species.
\end{abstract}

\section{Introduction}

\ce{H2} and \ce{H2O} are the most abundant gas phase and one of the most abundant solid phase molecular constituents in interstellar dense clouds, respectively. These particular astronomical objects are characterized by extremely harsh conditions, especially with respect to temperatures (10--50 K) \cite{Snow2006}. These low temperatures are due to the UV radiation filtering in the edges of the object by cosmic dust. Another consequence of this radiation filtering is that molecules can be formed and, more importantly, withstand the conditions of the region. Prevalence of molecular material at these temperatures lead to the formation of ice mantles on top of carbonaceous and siliceous dust grains by accretion of molecules.

\ce{H2} is a key molecule for the chemical evolution of a dense cloud both in its neutral form, but specially when ionized \cite{Dalgarno2000}. Its role in the astrophysical picture, however, is not limited to chemical reactions. Molecular hydrogen is also an effective coolant, and a molecular thermometer due to its nuclear spin properties \cite{Bron2016,Wakelam2017}. Ortho-para ratio (OPR) evolution models consider a return of the molecular hydrogen to the solid phase, adsorbing on interstellar ices, as a first step of the model \cite{Furuya2019}. In addition, gas-grain reactions are of paramount importance for astrochemistry, with dust grains acting as templates or reservoirs of molecules. It has been previously stated that, if no mechanism of return to the gas phase would be acting on dust grains, all molecules would remain stuck in interstellar dust grains \cite{Willacy1998}. The quantitative magnitude governing adsorption is the so-called \emph{sticking coefficient}, $S_\text{T}$, a  magnitude that encompasses all the physics of adsorption of a surface-adsorbate pair. Formally, the sticking coefficient is a quantity between 0 and 1 representing the fraction of molecules that remain stuck on a surface after a collision. Sticking coefficients are usually considered to be a function of three main factors, gas temperature, surface temperature and mass of the incoming particle. The reality is that many factors are at play, such as surface morphology, geometry of the collision, etc. Because of this complexity, and with the low temperature conditions of dense clouds, astrophysical models typically approximate sticking coefficients by constant values of either 0.5 or 1.0 depending on the molecule \cite{Aikawa2012, Chang2012}. Recent efforts in the literature are opposing this trend by providing values of this magnitude for different systems, setups and gas temperatures. These measured sticking coefficients are luckily being incorporated in models, increasing the complexity of the description of the physical and chemical magnitudes being calculated \cite{Furuya2019}.

The specific role of sticking coefficients in grain surface models has been reviewed recently \cite{Cuppen2017}. Briefly, sticking affects the rate of accretion ($f_\text{acc}$) of a molecule on a dust surface, weighting the collision frequency of a gas-phase adsorbate of abundance $n_\text{g}$ according to the following expression.

\begin{equation}
f_\text{acc}=S_\text{T} \nu n_\text{grain} \pi r^{2}_\text{grain}  n_\text{g}
\end{equation}

With $\nu$ being the average thermal velocity of the gas, $n_\text{grain}$ and $r_\text{grain}$ are abundance and average radius for the grain, respectively. The rate of accretion $f_\text{acc}$ constitutes the first term in gas-grain models, representing the landing of chemicals on the catalytic surface. Sticking can, thus, be considered as the first step for interstellar surface chemistry to occur.

Experimental and theoretical studies have addressed the determination of sticking coefficients of astrophysical molecules in different ways. In the experimental part, sticking coefficients have been determined for a variety of closed-shell light molecules and the hydrogen atom on different substrates varying molecular beam or surface temperature \cite{Hornekar2003,Matar2010,Chaabouni2012, Acharyya2014, He2016}. Despite some disagreements in the results, attributed to different experimental techniques or conditions, all these studies reveal that sticking coefficients are not a binary magnitude. Theoretically, two approaches to the determination of the sticking coefficient can be found, either the classical analytic solutions based on binding energies on simple models \cite{Burke1983, Jones1985, Leitch-Devlin1985} or by means of molecular dynamics simulations. Analytic models can not provide the required degree of accuracy and for molecular dynamics simulations and specifically in a \textit{force-field} context, the amount of studied systems remains small and far from general. \ce{H} has been extensively studied \cite{Buch1991,Masuda1998,Al-Halabi2007,Veeraghattam2014,Dupuy2016} and \ce{CO} has also received some attention \cite{Al-Halabi2003,Al-Halabi2004}. We have not been able to find data for \ce{H2}, apart from a PhD thesis chapter \cite{Veeraghattam2014thesis}. Addressing a different but similar topic, such as Eley--Rideal reactions to form \ce{H2}, Casolo et al \cite{Casolo2013} calculated reaction cross sections for the \ce{H}+\ce{H}$\rightarrow$ \ce{H2} reaction, using molecular dynamics simulations, this time in an \textit{ab-initio} flavor.

With this paper, we have two intentions. First, we will introduce a new general method for calculating sticking coefficients with \textit{quantum-chemistry } based molecular dynamics simulations making use of cheap electronic structure solvers capable of describing intermolecular interactions with sufficient accuracy. Second, we will employ our method to derive the sticking coefficients of the \ce{H2}/ASW system, in order to give a rational basis for the observations, discussing the limitations of both experiments and theory and how they can complement each other, providing valuable data to the astrophysical community.

\section{Methodology}

The present section is dedicated to the discussion of the theoretical tools employed in the work. It is divided as follows: First, a description of the construction procedure of the cluster mimicking a water ice surface will be given. Second, we will dedicate some effort to the description of the \ce{H2}/ASW sampling procedure in our method. Finally, we will rationalize the selection of the electronic energy and force solver, showing tests with model systems.

\subsection{\textit{Surface Construction}}

\subsubsection{Initial Set-up}
The method to prepare clusters to model the surface of amorphous solid water follows the same procedure as previous works of the group \cite{Song2016,Lamberts2017} with some modifications required for molecular dynamics runs. The starting water sample is generated using VMD \cite{VMD} version 1.9.2 containing 18937 TIP3P water molecules \cite{Jorgensen1981,CHARMM}. These molecules are packed in a periodic cubic cell of $\sim 80 \times 80 \times 80$ \angstrom$^3$, yielding a density of 1~g~cm$^{-3}$. The system was equilibrated at 200 K for 100 ps and then suddenly quenched to 10 K for another 10 ps. These simulations were carried out in the NVT ensemble, using a Langevin thermostat and the NAMD code \cite{NAMD}. From this initial system we extracted a hemispherical cluster of diameter 40 \angstrom. The center of the hemisphere is set as the origin of coordinates for further calculations.

\subsubsection{\textit{Preparation of the QM/MM System}}

Production runs on the system have been done employing multiscale embedding procedures, specifically the electrostatically embedded QM/MM \cite{QMMM} partition of the total system as implemented in the ChemShell suite of programs \cite{She2003,Chemshell}. The TIP3P model was used for the MM part. The choice of the QM methods requires some care and will be discussed below. The electrostatic embedding means that MM charges are included in the QM Hamiltonian. Dispersion interactions between QM and MM, as well as within MM atoms, are treated by the force field. Dispersion interaction parameters for the \ce{H2} molecule are taken from the literature \cite{Du2011}.

\begin{figure*}[h]
\begin{center}
\includegraphics[width=0.8\textwidth]{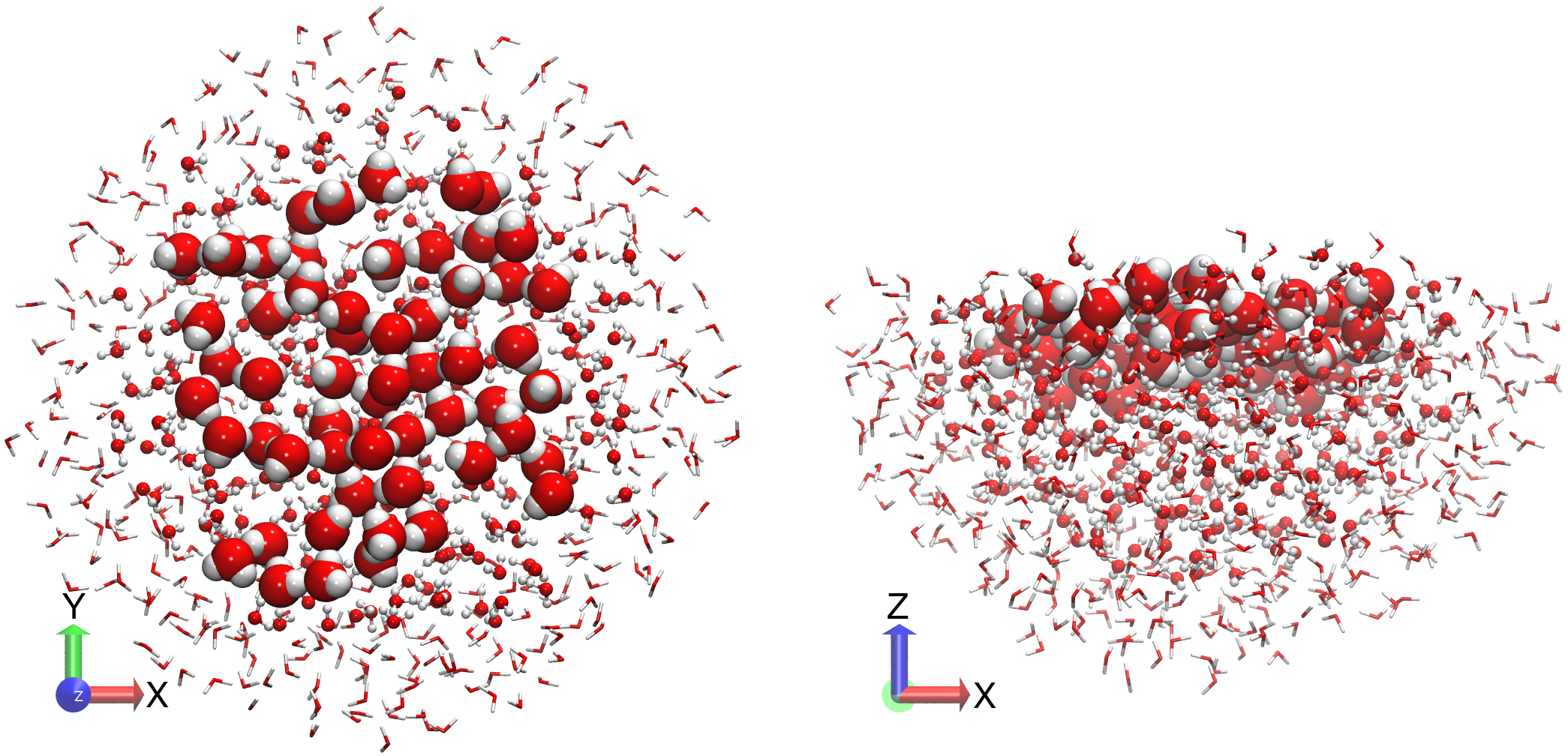}
\caption{Top (left) and side (right) views of the water hemisphere employed as a surface model for MD runs. QM atoms are displayed as large spheres, flexible MM atoms as balls and sticks, frozen MM atoms as sticks.}
\label{slabopt}
\end{center}
\end{figure*}

The QM part of our system needs to be large enough to allow \ce{H2} molecules enough room for diffusion after collision. Taking this into account, we selected all atoms within $20\times 20\times 3\ \angstrom^3$, centered at the center of the hemisphere to be treated at the quantum level (57 water molecules). These are indicated in \figref{slabopt} by large spheres. The remaining molecules are treated at the MM level (626 water molecules). All water molecules, QM and MM were kept internally rigid by means of the SHAKE algorithm, unless noted otherwise.\cite{SHAKE} We compared the effect of rigid and internally flexible water molecules and found it negligible, as described in the Discussion. The QM water molecules and the closest 235 MM water molecules to the QM part are allowed to move during the dynamics. These are indicated by a ball-and-stick representation in \figref{slabopt}. The rest of water molecules are kept frozen to provide an outer boundary during the MD runs. Improvement of the water model for the MM part is possible and currently being tested.

The QM/MM system is equilibrated following an initial geometry optimization in DL-Find \cite{DLFIND}. The short equilibration runs were performed in an NVT ensemble for 10~ps starting from random initial velocities. Systems were equilibrated at 10, 30, 70 and 110~K in order to calculate sticking coefficients at these temperatures. The temperature was kept fixed by a Nos\'e--Hoover thermostat. 20 snapshots were taken for each temperature. All the QM/MM dynamics (snapshots generation and trajectories) have been driven by the DL\_Poly suite \cite{DLPOLY} within ChemShell.

\subsection{\textit{Sampling Procedure}} \label{sampling_strategy}

With the chemical model for the surface prepared, we are finally in position to explain the strategy to sample collision dynamics between adsorbate (\ce{H2} in this particular case, but general for every molecule) and the surface. The philosophy behind the method is to guarantee that the adsorbate (also treated at the quantum level within the QM/MM framework) hits always in the quantum part of the surface. For doing so, the strategy is to always define the impact point on the surface as the center of the previously mentioned quadratic selection within the hemisphere plus a small random displacement in the $X$ and $Y$ direction, being the $Z$ direction normal to the surface plane. We have chosen a maximum of 4 \angstrom{} for the random displacement. At higher values the adsorbate, slightly attracted from the surface, could collide with the edges of the cluster for some configurations. This displacement is always lower than the dimensions of the QM selection to ensure sampling of the surface plane. This, in combination with the use of different MD snapshots as initial configurations for the surface, guarantees that for every trajectory we have a different collision point, as expected from an amorphous surface. 

\begin{figure}[h]
\begin{center}
\includegraphics[width=0.50\textwidth]{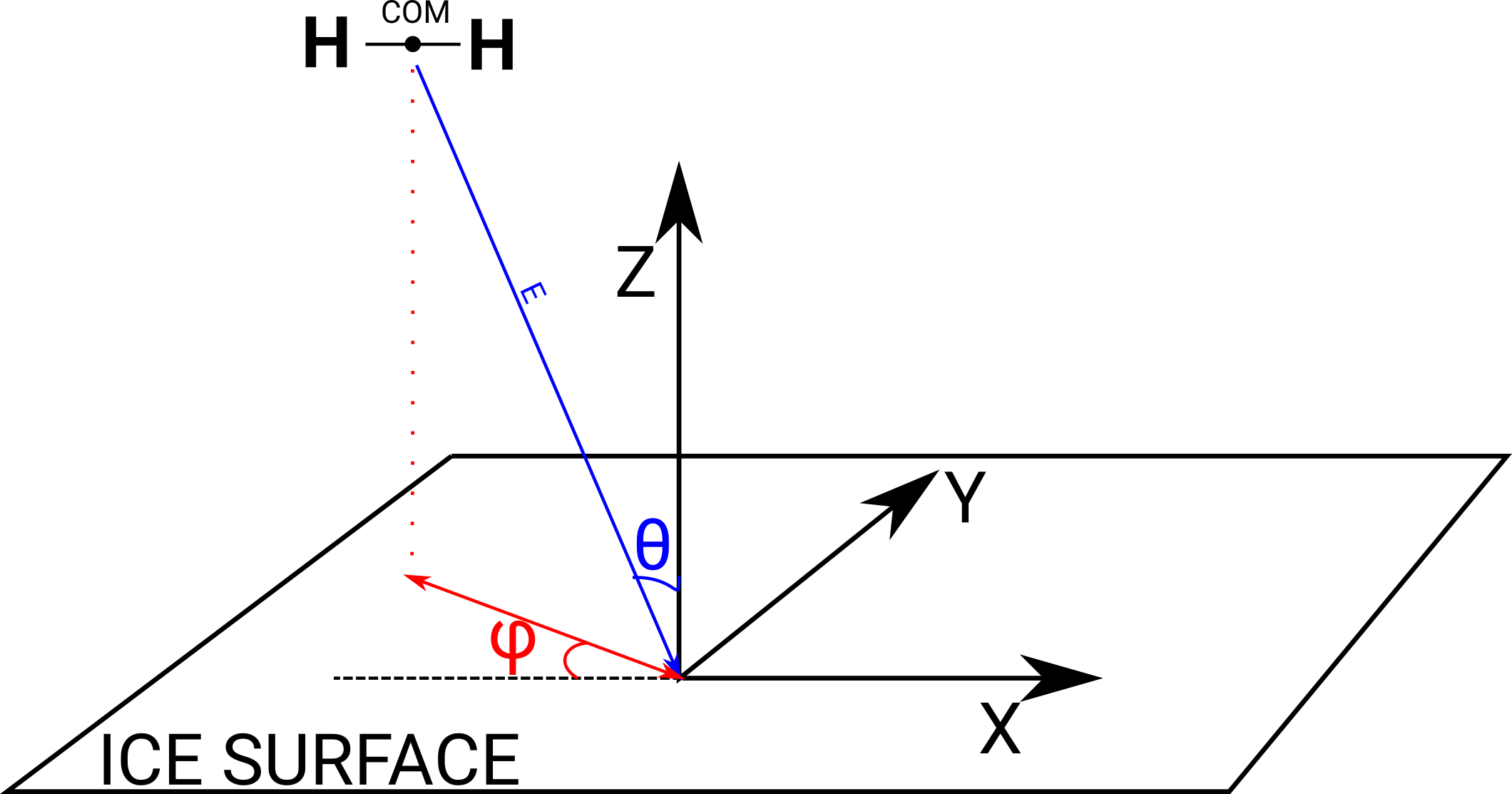}
\caption{Scheme of the sampling strategy employed in our method. The initialization parameters are the incident energy ($E_{i}$), the polar angle of incidence ($\theta$), and the azimuthal angle of incidence ($\phi$). Other initialization parameters are the random lateral displacement in the $XY$ plane of the surface and the random initial orientation of the molecule.}
\label{Referenciasamp}
\end{center}
\end{figure}

With the theoretical collision point defined, we can calculate the velocity vector from the center of mass of the adsorbate. The polar $(\theta)$ angle of incidence with respect the surface normal is sampled from a distribution of angles with a theoretical maximum value of 90\degree, for a tangential incidence. The reality is, given the use of a cluster model, the module of the velocity vector impinging at 90{\degree} would be infinite. In practice, we define an upper value for the $(\theta)$ angle, to avoid distant starting points for the adsorbate. See \figref{Referenciasamp} for a schematic view of our sampling strategy. We have used 65{\degree} as the maximum $\theta$ value for all of our simulations. The number of different $\theta$-values is an input parameter of the model. Unless stated otherwise we have employed 14 different $\theta$ values in our simulations. For each value of $\theta$ we have sampled different values of the azimuthal angle ($\phi$). To have an equally sampled solid angle, the number of $\phi$-values used follows the relation:
\begin{equation} \label{solid}
\phi_\text{used}=\max\left[1,\text{int}\left(2\phi_\text{max}\sin(\theta)\cos(\theta)\right)\right]
\end{equation}
where $\phi_\text{max}$ is a user defined value, also chosen to be 14 as a default. The two remaining input parameters are the minimum initial $Z$ component of the center of mass of the adsorbate with respect to the surface (8 \angstrom) and the incoming velocity of the adsorbate's center of mass. In our model, all the kinetic energy of the adsorbate is translational, corresponding to a particle in its ro-vibrational ground state in a quantum mechanical picture. We randomize the initial orientation of the adsorbate. However, we found that the initial orientation of the adsorbate only plays a minor role since it slowly aligns with the surface according to its experienced potential. A sweep in translational energies ($E$) between 1 and 60 meV for every surface temperature has been performed. 

The trajectory is obtained by integrating Newton's classical equations of motion for the nuclei with the velocity-Verlet algorithm \cite{Verlet1969} using a time step of 0.5~fs. This resulted in a sufficient energy conservation. The sampling of the collision is performed in an NVE ensemble, because a thermostat would artificially remove energy from the adsorbate before the impact, so that the collision energy would deviate from the desired one.

Each simulation is run until a bouncing event happens or the maximum simulation time is reached. Veeraghattam et al. \cite{Veeraghattam2014, Veeraghattam2014thesis} found in force-field MD simulation of H sticking in ASW no dependence of the sticking coefficients on the total simulation time, suggesting 5 ps. In most cases, we have followed that suggestion. However for surface temperatures $>10$ K (in which surface-adsorbate energy transfer becomes important), we have increased the simulation time as indicated. 

For each trajectory, the analysis must distinguish between bouncing and sticking events. Dupuy et al. \cite{Dupuy2016} suggested a criterion for force-field MD based on the potential energy of the incoming molecule. Such a scheme is not applicable in quantum-chemistry based MD, since the potential energy cannot be partitioned onto individual atoms. Because of that, we cannot automatize the termination of the dynamics in the case sticking events. Bouncing events, i.e. simulations with the adsorbate returning to the gas phase, are identified by the minimum distance between the adsorbate and any other atom on the surface becoming larger than 7 \angstrom. Any other trajectories are considered as a sticking event. Visual inspection of many trajectories confirms this assignment. The ratio of sticking vs total number of trajectories is the \textit{sticking probability} $P(E)$. Bouncing and sticking events follow a binomial distribution, which we use to provide the statistical uncertainty using the Jeffreys interval \cite{Brown2001}, as implemented in the Scipy mathematical library \cite{Scipy}. 
 
In total, for a run with 14 values for $\theta$ and $\phi_\text{max}=14$, we run a total of 138 trajectories per incoming energy, with a total of 14 incoming energies per surface temperature.

\subsection{\textit{Calculation of Sticking Coefficients\label{calcst}}}

Sticking probabilities $P(E)$ are obtained as the ratio between the number of trajectories that lead to adsorption and the total number of simulation runs. From the set of obtained values for $P(E)$ at different incident energies we obtained a thermally averaged sticking coefficients ($S_\text{T}$)  performing a Maxwellian integration of the type \cite{Buch1991, Veeraghattam2014thesis}:
\begin{equation} \label{thermalaveraged}
S_\text{T}=\dfrac{1}{(kT)^{2}}\int^{\infty}_{0}P(E)Ee^{-\frac{E}{kT}}dE
\end{equation}
where $E$ represents the adsorbate incoming energies and $k$ and $T$ are the Boltzmann constant and the temperature of the gas, respectively. We have employed two ways of integrating \eqref{thermalaveraged}. The first one is to linearly interpolate the values of $P(E)$, followed by integration. As boundary conditions we have employed $P(\text{0 } \text{meV})$ = $P(\text{1 } \text{meV})$ and $P(\infty)$= 0 at a given surface temperature. The second one is the approximation of $P(E)$ by the expression proposed by Buch \& Zhang \cite{Buch1991}, assuming an exponential dependence for $P(E)$ \cite{Buch1991}:
\begin{equation} \label{buch}
P(E)=e^{-\frac{E}{E_{0}}}.
\end{equation}
$E_{0}$ is a fitted parameter. The expression for the latter case (labeled as Buch \& Zhang in the graphs) then becomes:
\begin{equation}
  S_\text{T}=\left[\left(\frac{kT}{E_{0}}\right)+1\right]^{-2}
  \label{buchint}
\end{equation}

\subsection{\textit{Selection of the Quantum Chemical Method}} \label{sectionselection}

For the derivation of a single value of $P(E)$ we require on the order of 10$^{2}$ trajectories, each one consisting of around 10$^{4}$ dynamics steps, which results in a rough estimation of about 10$^{6}$ energy and gradient evaluations per $P(E)$ value. On top of that, several incident energies are required in order to sample a statistical distribution of velocities. Conventional quantum chemistry methods, namely density functional theory or wave-function based methods, are simply too computationally expensive to be of use here. Modern semi-empirical methods, and specially tight-binding ones, have been considered. In the particular case of the \ce{H2}/\ce{H2O} system, proper description of van der Waals interactions is required. We have selected the recently published GFN-xTB method \cite{Grimme2017}, in its second parameterization (GFN2-xTB) \cite{Bannwarth2019} as QM method for the QM/MM setting in  our quantum-chemistry based MD set-up. This method has the very recent D4 London dispersion model \cite{Caldeweyher2019} incorporated, which shows a higher degree of accuracy than the previous D3 model \cite{Grimme2010}, currently a standard in conventional DFT methods. We have implemented an interface between the GFN2-xTB code and ChemShell to employ the method in a QM/MM framework. 

\begin{figure}[h]
\begin{center}
\includegraphics[width=0.5\textwidth]{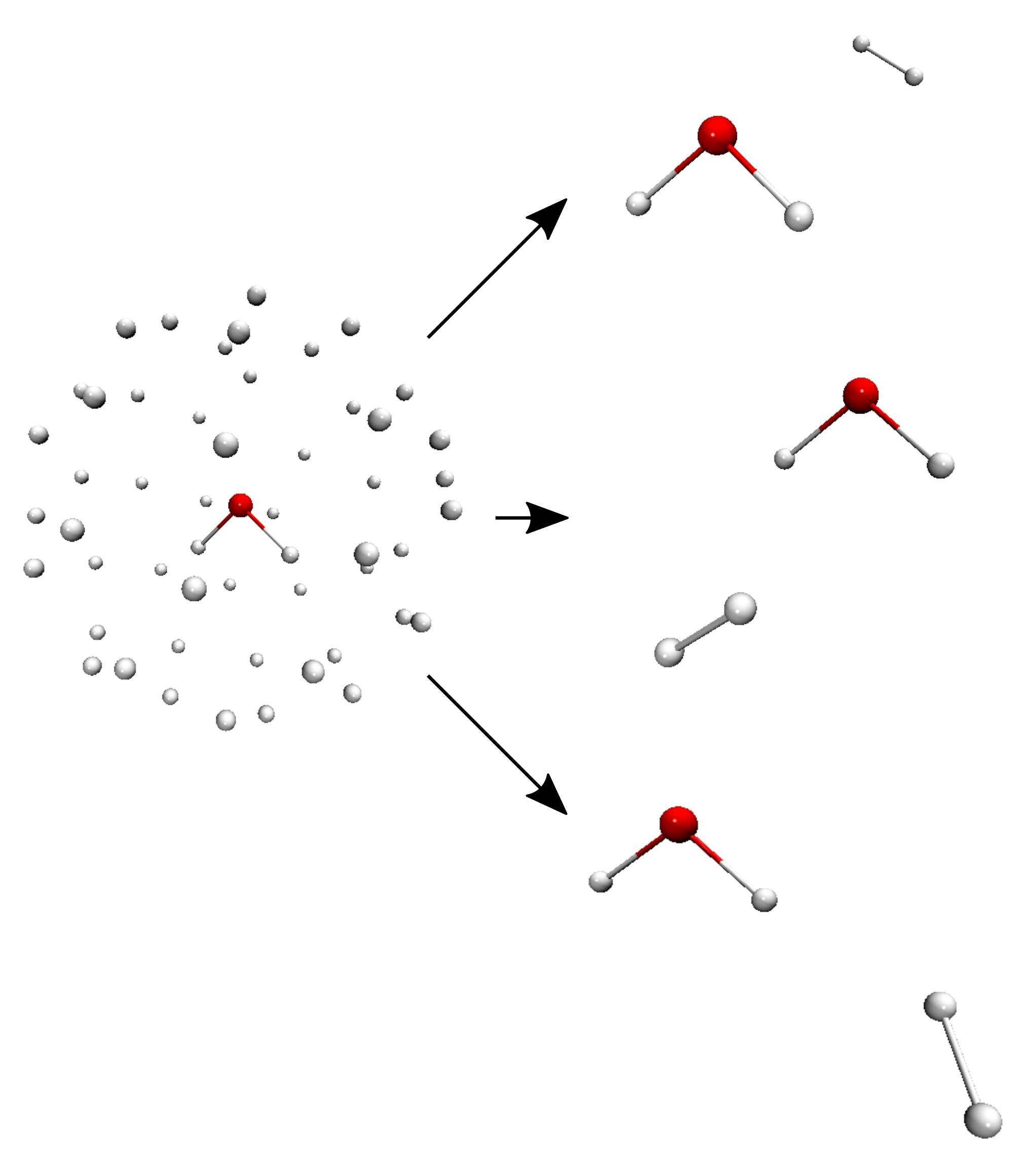}
\caption{Sequential steps for the construction of the gas phase model for benchmark purposes. I) Generation of an evenly distributed enclosing sphere of hydrogen atoms. II) Placing of the second hydrogen atom for every individual configuration. The shown sphere radius is 2.5 \angstrom. }
\label{esquema_bench}
\end{center}
\end{figure}

The quality of the interaction energies between H$_2$ and water was tested for a large number of configurations in a small gas phase model system comprising one water molecule and one hydrogen molecule. The procedure we have followed is to construct a sphere of evenly spaced points around the center of mass of the water molecule, on which we have placed one hydrogen atom. After that, a second hydrogen atom has been attached to the previous one, with random orientation, defining a configuration. We have done this for 8 sphere radii (1.5, 2.0, 2.5, 3.0, 3.5, 4.0, 5.0 \& 6.0 \angstrom ), with 50 configurations per radius. See \figref{esquema_bench} for a schematic view of the construction procedure. For each configuration we have tested the values of the interaction energy of GFN2-xTB against results at the B3LYP-D3BJ/def2-TZVP and the CCSD(T)-F12/VTZ-F12 levels. We have employed Turbomole (version 7.0.1) \cite{Turbomole} and Molpro (2015) \cite{MOLPRO} codes for these calculations, respectively. Interaction energies have been computed as $\Delta E_\text{int}= E_\text{con}-(E_{\ce{H2O}}+E_{\ce{H2}})$. $E_\text{con}$ represents the energy of a configuration whereas $E_{\ce{H2O}}$ and $E_{\ce{H2}}$ represent the water and hydrogen energies.

\begin{table}[h]
\centering
\caption{MAE of the binding energies of GFN2-xTB and B3LYP-D3BJ vs the reference CCSD(T)-F12 method as a function of the radii of the sphere of configurations}
\label{Benchmartab}
\begin{tabular}{ccc}
\hline
Radius Sphere  & MAE$_\text{B3LYP-D3BJ}$ &  MAE$_\text{GFN2-xTB}$ \\
(\angstrom) & (kcal/mol) & (kcal/mol) \\
\hline
1.5 & 1.877 & 11.126 \\
2.0 & 0.552 & 3.318 \\
2.5 & 0.179 & 0.755 \\
3.0 & 0.087 & 0.192 \\
3.5 & 0.060 & 0.081 \\
4.0 & 0.026 & 0.028 \\
5.0 & 0.011 & 0.014 \\
6.0 & 0.006 & 0.005 \\
\hline
\end{tabular}
\end{table}

\begin{figure}[h]
\begin{center}
\includegraphics[width=0.50\textwidth]{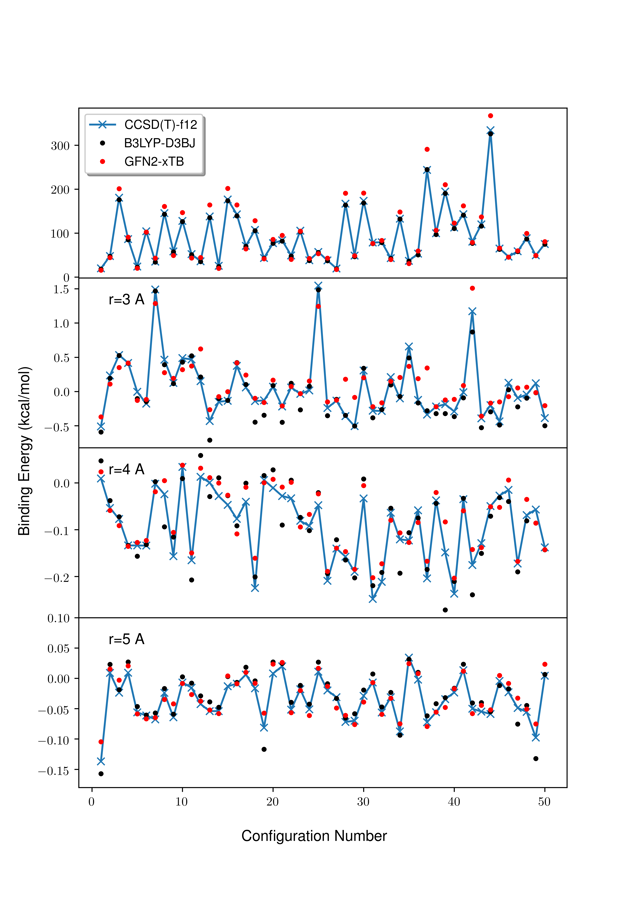}
\caption{Benchmark of the distribution interaction energies for the \ce{H2O}/\ce{H2} system for different radii of enclosing sphere. Lines for the CCSD(T)-f12 calculations are a guide to the eye.}
\label{Benchmarkimg}
\end{center}
\end{figure}

\figref{Benchmarkimg} shows the distribution of the interaction energies for selected spheres in the intermediate range. \tabref{Benchmartab} shows the the mean absolute error (MAE) in the calculated interaction energies between CCSD(T)-F12 and the other two methods. Overall, the agreement of GFN2-xTB with CCSD(T)-F12 is satisfactory over the relevant interaction range.  At radii relevant for the dynamics, the error is below the chemical accuracy of 1 kcal/mol. Only if H$_2$ approaches the water molecule very closely, in the strongly repulsive areas at low radius, the error increases above that threshold. Such close encounters will, if at all, only be sampled at high incident energies. We have therefore confirmed that GFN2-xTB is a robust and adequate method to treat this system using it for the rest of the work. \tabref{tablebindingmonomer} shows the average interaction energy per radius of the sphere for the three methods under consideration. From the table we can extract that interaction energies at the DFT and semi-empirical levels reproduce the more exact values of the coupled-cluster calculations quite well. We plan to use this spherical sampling approach to determine binding energies for a variety of relevant adsorbates on water cluster models (dimer, trimer, tetramer) in further studies.
 
\begin{table}[h]
\centering
\caption{Average interaction energy (in kcal/mol) of the \ce{H2}/\ce{H2O} system as a function of the sampling sphere radii.}
\label{tablebindingmonomer}
\begin{tabular}{cccc}
\hline
Radius Sphere  & \multicolumn{3}{c}{$\Delta E_\text{int}$} \\
$(\angstrom)$ & CCSD(T)-F12 & B3LYP-D3BJ & GFN2-xTB\\
\hline
1.5 & 92.29  & 90.44 & 99.45 \\
2.0 & 20.40 & 19.89 & 21.76 \\
2.5 & 2.99 & 2.86 & 3.25 \\
3.0 & 0.05 & $-$0.01 & 0.12 \\
3.5 & $-$0.12 & $-$0.16 & $-$0.01\\
4.0 & $-$0.09 & $-$0.09 & $-$0.08 \\
5.0 &  $-$0.03 &  $-$0.03 & $-$0.03\\
6.0 &  $-$0.01 &  $-$0.01 & $-$0.01\\
\hline
\end{tabular}
\end{table} 

\section{Results}

\subsection{\textit{Binding Energies of \ce{H2} on ASW}}

The physics of an adsorption event are easy to rationalize. A particle is adsorbed on top of a surface when it is trapped in a potential energy well of a certain depth. What determines if an arbitrary event leads to adsorption or ejection to the media is the binding energy of the adsorbate just after the collision. After a collision, if the remaining kinetic energy of the adsorbate is enough to escape the potential well, the molecule will move either to another potential well or to the media, the latter producing a bouncing event. In systems dominated by dispersion forces, as the one we are handling here, binding energies are expected to be rather small and thus, the molecule is likely to visit several small wells before remaining bound in one.

\begin{figure}[h]
\begin{center}
\includegraphics[width=0.5\textwidth]{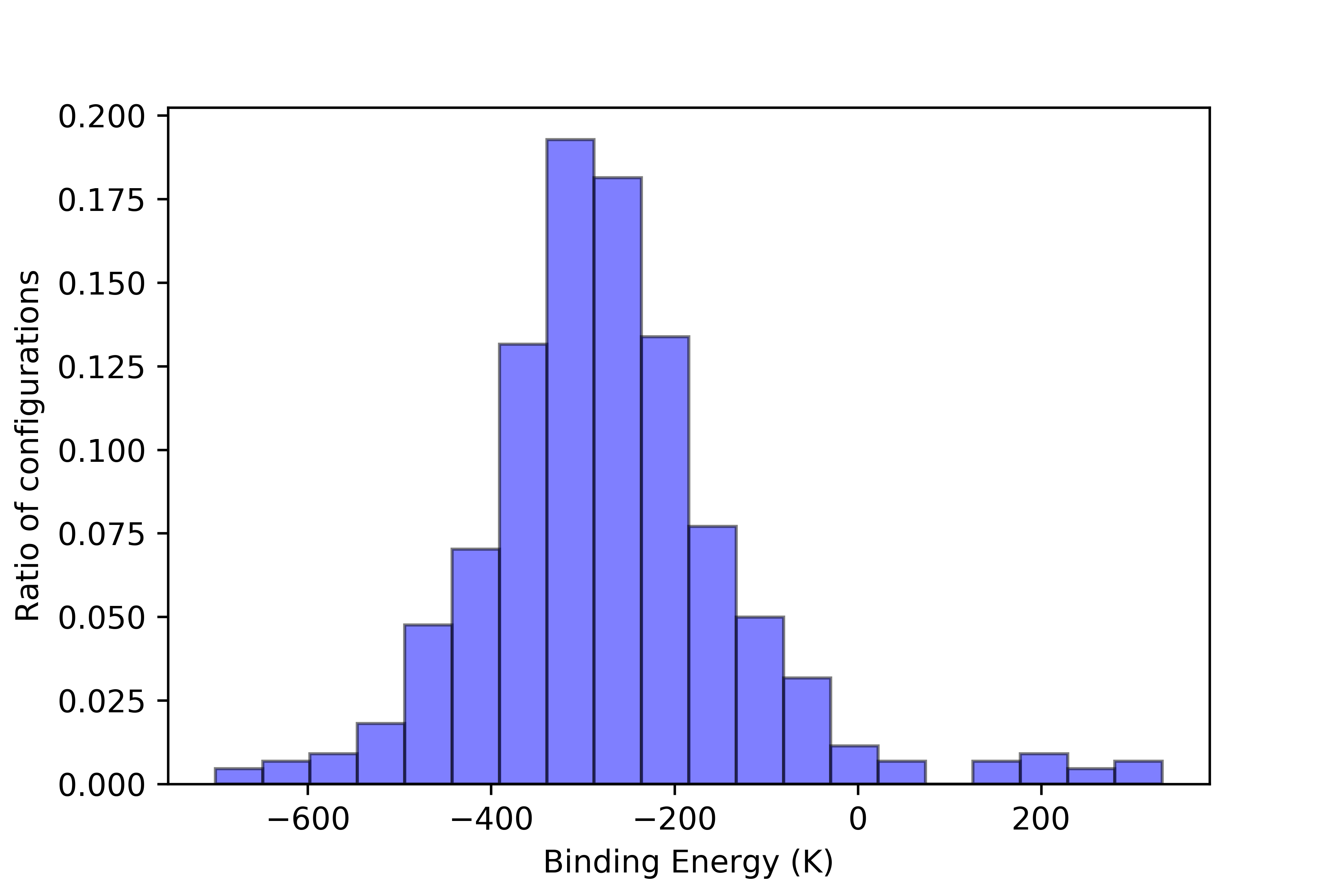}
\caption{Distribution of binding energies (negative values represent attractive configurations and positive repulsive ones) for the \ce{H2}/ASW system.}
\label{binding_fig}
\end{center}
\end{figure}

We have calculated the binding energies of \ce{H2}/ASW in our system in order to see the magnitude of the potential wells. The procedure to obtain the distribution of binding energies has been made placing the \ce{H2} molecule normal to the surface at a constant height of 3.0 $\angstrom{}$ and moving the molecule by steps of 1 $\angstrom{}$ in the $X$ and $Y$ direction spanning the limits of the QM part ($20\times 20$ $\angstrom^2$). For every initial position of the \ce{H2} molecule we have carried out a geometry optimization. This gives us a set of 400 binding energies. We have calculated the binding energy by $\Delta E_\text{bin}= E_\text{con}-(E_{\ce{ASW}}+E_{\ce{H2}})$. Results for the distribution of binding energies can be seen in \figref{binding_fig}. Around 75\% of the conformations fall in the range between 450--150 K, corresponding to weakly bounded physisorbed states. The values given are potential energies. The vibrational zero point energies would shift the values to somewhat weaker binding. Results for the binding energy are slightly lower than the ones presented for atomic hydrogen \cite{Al-Halabi2007}, that has an average value of 600 K. Our results are also in agreement with experimental values of the literature (see Table 1 of Bovino et al. 2017, \cite{Bovino2017}) It is important also to remark that some configurations are slightly repulsive, due to steric interactions between \ce{H2} and a hydrogen from a water molecule in the surface.

\subsection{\textit{Dependence of the Sticking Coefficients on the Incident Energy}}

As stated in the methodology section, several factors have been studied in our simulations. We will start by addressing the influence of the kinetic energy of the incoming \ce{H2} molecule at constant surface temperature. We will use the results at a surface temperature of 10 K for this section. \tabref{conditionsenergy} shows the conditions for the results that will be presented hereafter. We highlight the use of total simulation time of 5 ps. For 10 K, in our tests (considering also 30 ps), we have found 5~ps to be sufficient.

\begin{table}[h]
\centering
\caption{General conditions for the MD productions at a surface temperature of 10 K.}
\label{conditionsenergy}
\begin{tabular}{ccc}
Sampled energies  & 1,3,5,8,10,15,20,25 \\
(meV)  & 30,35,40,45,50,60 \\
Max random lateral displacement (\angstrom) & 4  \\
N$_{\theta}$ & 14  \\
$\phi_\text{max}$ & 14  \\
Thermodynamic ensemble & NVE  \\
Initial height with respect to the surface ($\angstrom$) & 8  \\
Maximum time of the dynamics (ps) & 5  \\
Integration time step (fs) & 0.5  \\
Total number of trajectories & 1932 \\
\end{tabular}
\end{table}

\begin{figure*}[h]
\begin{center}
\includegraphics[width=0.80\textwidth]{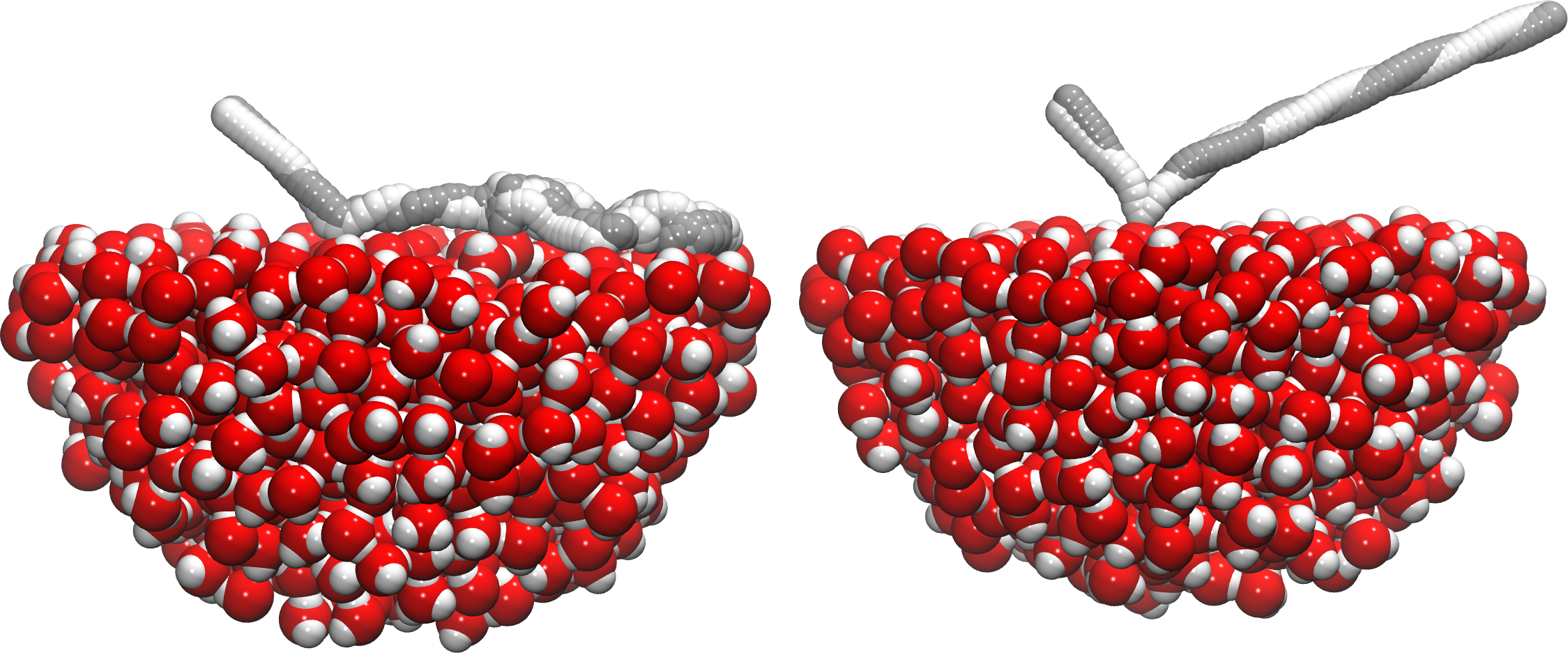}
\caption{Representative trajectories for sticking (left) and bouncing (right)
  events at an incident energy of 25 meV and a surface temperature of 10~K. The two atoms of H$_2$ are colored
  white and gray.}
\label{trajectories}
\end{center}
\end{figure*}

A static representation of two trajectories is given in \figref{trajectories}. For both, we can unambiguously determine the nature of the collision event, categorizing them respectively as sticking or bouncing trajectories. We have followed a similar procedure for every one of the 1932 trajectories considered in this section. The criterion of labeling a trajectory as bounced if the minimum distance to the surface with respect to any other atom is higher than 7 $\angstrom{}$ has proven simple yet powerful for all the trajectories we have visually inspected. From the images we can see furthermore that before collision, some rotational movement due to the slight attractive interaction potential can be attributed to the \ce{H2} molecule. After collision, an important part of the translational energy is the rotational motion. From a fundamental point of view, transfer of energy from the adsorbate to the surface and redistribution of the initial energy in different degrees of freedom are the main factors behind a sticking event.

\begin{figure}[h]
\begin{center}
\includegraphics[width=0.50\textwidth]{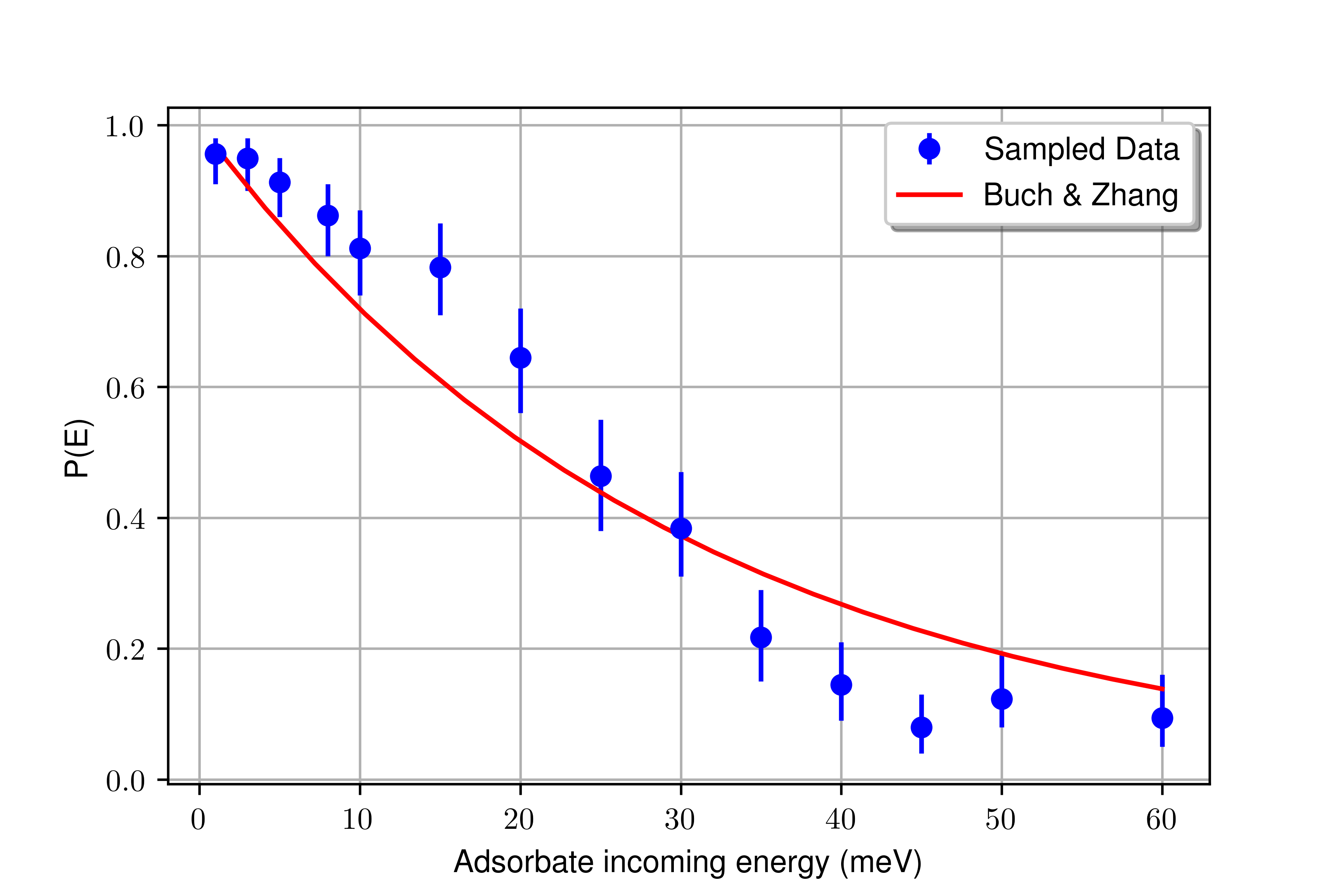}
\caption{Theoretical sticking probabilities for different adsorbate incoming energies for a surface temperature of 10 K. In blue discrete points: Values from our simulation. Red continuous line: Buch \& Zhang fit, \eqref{buch}. 95\% confidence intervals are indicated.}
\label{sticking_energies}
\end{center}
\end{figure}

The results for the sticking probability $P(E)$ for each incoming energy are shown in \figref{sticking_energies} and \tabref{sticking_energies_tab}. From the graph, several facts can be inferred. We can observe that at low incoming energies, up to 10 meV ($\sim$ 115 K), the values of $P(E)$ remain close to unity. At higher incoming energies, in the intermediate part of the graph, the sticking probability steeply decreases. At even higher incoming energies we also find a plateau (between 40 and 60 meV), implying a saturation in the bouncing events at these energies. It is possible to further decrease the sticking probability to almost zero, at much higher incoming energies. One can see clear fluctuations in the values, which are explained by the statistical error of the sampling approach. These fluctuations are averaged out in the thermal integration. We also present the values of $P(E)$ using the proposed formula by Buch \& Zhang. Values for $E_{0}$ are obtained from the least-squares fit of \eqref{buch}, which resulted in  $E_{0} = 30.3$ meV. 
 It can be observed that an exponential decrease is not adequate for the description of $P(E)$ for this system. Especially the slow decay trend for small $E$ is incompatible with an exponential behavior.

\begin{table}[h]
\centering
\caption{Values of the sticking probabilities at different incident incoming
  energies for the adsorption of \ce{H2} on ASW at 10 K. The 95\% confidence
  interval is provided in parentheses.}
\label{sticking_energies_tab}
\begin{tabular}{cc}
\hline
$E$ (meV) & $P(E)$\\
\hline
1 & 0.95 (0.91--0.98)   \\
3 & 0.94 (0.90--0.98)   \\
5 & 0.91 (0.86--0.95)   \\
8 & 0.86 (0.80--0.91)   \\
10 & 0.81 (0.74--0.87)   \\
15 & 0.78   (0.71--0.85)   \\
20 & 0.64  (0.56--0.72)  \\
25 & 0.46  (0.38--0.55)  \\
30 & 0.38  (0.31--0.47)  \\
35 & 0.22  (0.15--0.29)  \\
40 & 0.14  (0.09--0.21)  \\
45 & 0.08  (0.04--0.13)  \\
50 & 0.12  (0.08--0.19)  \\
60 & 0.09  (0.05--0.15)  \\ 
\hline 
\end{tabular}
\end{table}

\begin{figure}[h]
\begin{center}
\includegraphics[width=0.50\textwidth]{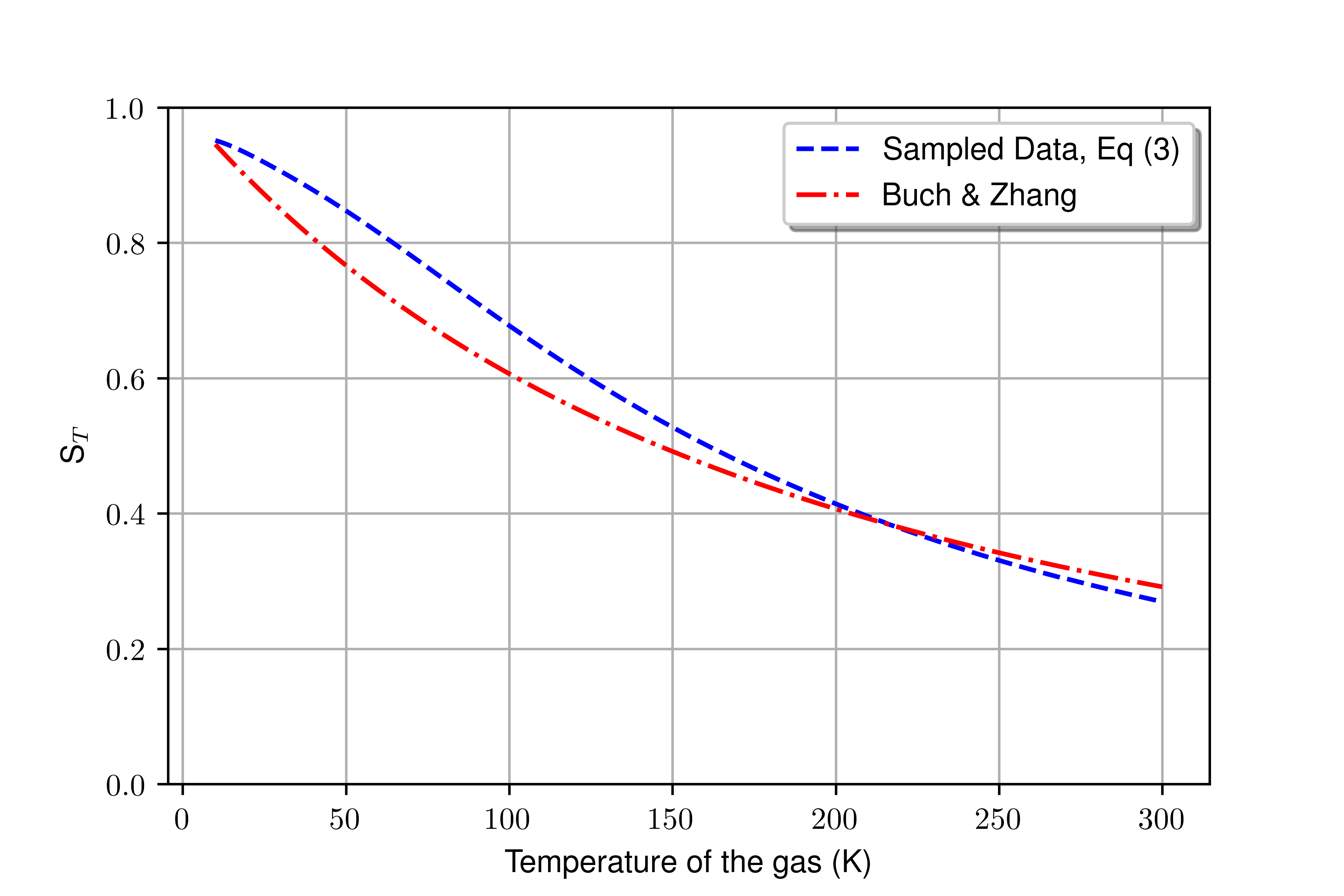}
\caption{Theoretical sticking coefficients at different gas temperatures. In red \eqref{thermalaveraged} and in blue \eqref{buchint}. Temperature of the surface is 10 K}
\label{sticking_coefficients}
\end{center}
\end{figure}

The resulting thermal sticking coefficients can be found in \figref{sticking_coefficients} and \tabref{sticking_coefficients_tab}.  The obtained results with a linear interpolation of $P(E)$ show that even at a gas temperature of 10 K, the sticking coefficient of \ce{H2} is not unity, but is it is close (0.95).  At higher temperatures of the gas (300 K), for a surface temperature of 10 K we find a value of $S_\text{300\text{ K}}=0.27$.  The results using the Buch \& Zhang fit from \eqref{buchint} underestimate the sticking coefficient at low gas temperatures. 

\begin{table}[h]
\centering
\caption{Values of the sticking coefficients as a function of the gas temperature for the adsorption of \ce{H2} on ASW at 10 K.}
\label{sticking_coefficients_tab}
\begin{tabular}{ccccc}
\hline
$T_\text{gas}$ (K) & $S_\text{T}$   \\
\hline
10 & 0.95   \\
30 & 0.90    \\
50 & 0.85    \\
100 & 0.68    \\
120 & 0.62    \\
150 & 0.53    \\
180 & 0.46   \\
200 & 0.41   \\
230 & 0.36   \\
250 & 0.33   \\
280 & 0.29   \\
300 & 0.27   \\
\hline 
\end{tabular}
\end{table}

\subsection{\textit{Dependence of the Sticking Coefficient on the Angle of Incidence}}

\begin{figure}[h]
\begin{center}
\includegraphics[width=0.50\textwidth]{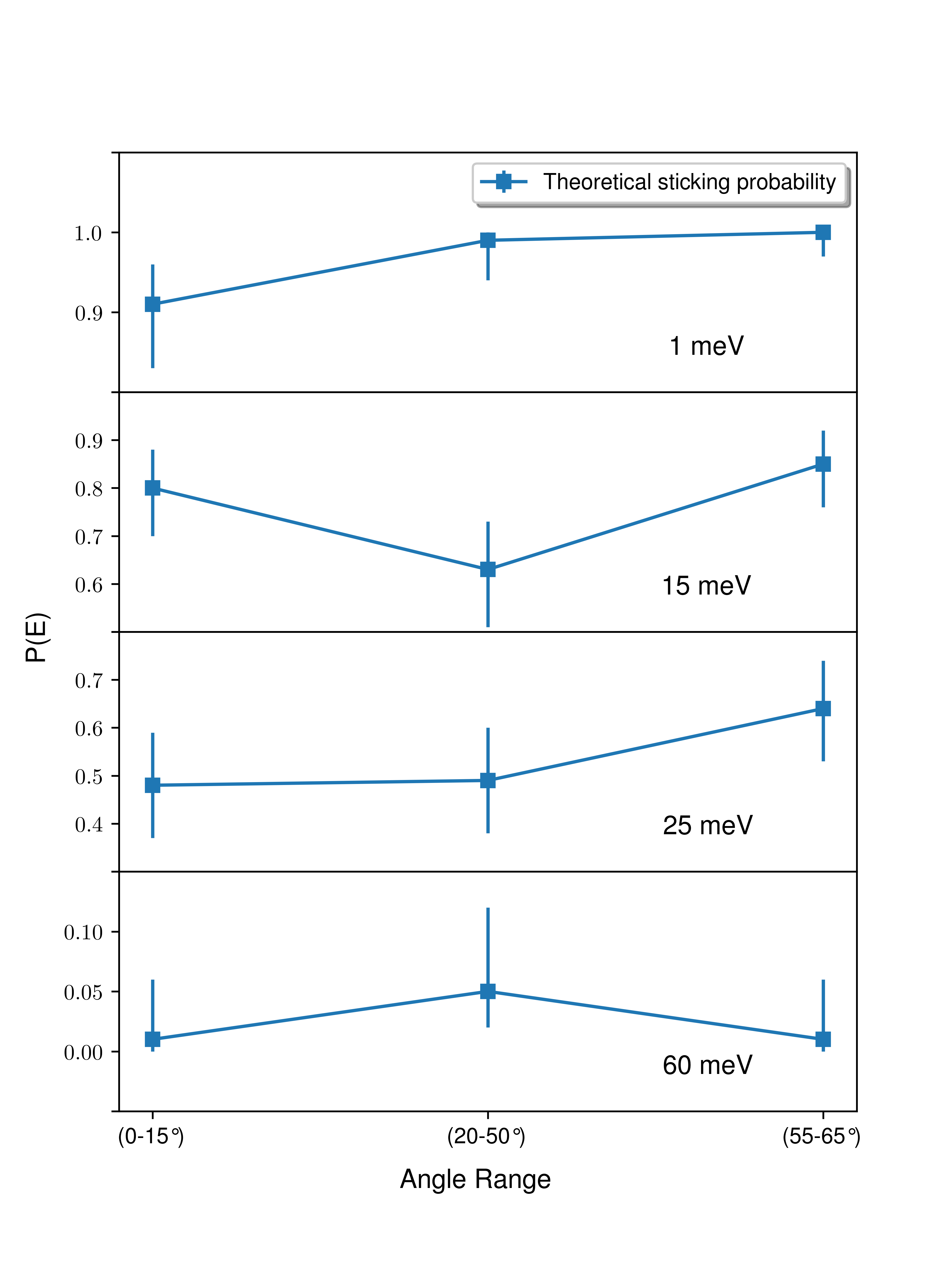}
\caption{Theoretical sticking probabilities as a function of the angle of incidence ($\theta$) for the \ce{H2}/ASW system at 10 K.}
\label{angles}
\end{center}
\end{figure}

For the dependence with respect to the angle of incidence $\theta$ we performed individual evaluations of $P(E)$ at a surface temperature of 10 K for four initial adsorbate velocities (1, 15, 25 \& 60 meV). For each of these velocities we used three different intervals of angular sampling, namely \textit{quasinormal} (0--15\degree), \textit{intermediate} (20--50\degree) and \textit{grazing} (55--65\degree). All angles were sampled equally, independently of their contribution to the solid angle. That is why we present the results of $P(E)$ instead of performing a Maxwellian integration. We sampled, for each interval, a total of 5 equi-spaced $\theta$-values and 15 $\phi$-values. \figref{angles} and \tabref{tabangles} show the results on the influence of the angle.

\begin{table}[h]
\centering
\caption{Values of the angle dependence of the sticking probabilities at discrete incident incoming energies for the adsorption of \ce{H2} on ASW at 10 K including the 95\% confidence interval.}
\label{tabangles}
\begin{tabular}{ccc}
\hline
$E$ (meV) & $\theta$ (deg.) & $P(E)$  \\
\hline
1 & 0--15 &  0.91 (0.83--0.96)   \\
1 & 20--50 &  0.99 (0.94--1.00)  \\
1 & 55--65 &  1.00 (0.97--1.00)   \\
\hline
15 & 0--15 &  0.80 (0.70--0.88)   \\
15 & 20--50 &  0.63 (0.51--0.73)    \\
15 & 55--65 &  0.85 (0.76--0.92)   \\
\hline
25 & 0--15 & 0.48 (0.37--0.59) \\
25 & 20--50 & 0.49 (0.38--0.60) \\
25 & 55--65 & 0.64 (0.53--0.74)\\
\hline
60 & 0--15 & 0.01 (0.00--0.06)\\ 
60 & 20--50 & 0.05 (0.02--0.12)\\
60 & 55--65 & 0.01 (0.00--0.06)\\ 
\hline 
\end{tabular}
\end{table}

We observe that for the quasinormal and intermediate angles of incidence no statistically significant conclusions can be drawn on the angle dependence of the sticking probability. However, for grazing angles, a slight increase in the sticking probability is observed in all plots except of 60~meV. The latter can be explained by saturation of bouncing events, see above. There is an indication of a relation between the sticking probability and the angle of incidence. The difference between grazing incidence and lower angles increases at intermediate incoming energies (see values in \tabref{tabangles}).
It is important to remark that the overall influence of the angle of incidence is rather low. It constitutes about 3--9\% at low and high incident energies and about 15\% at intermediate incoming energies. However, the uncertainty in the determination of $P(E)$ is higher for intermediate incoming energies.

\subsection{\textit{Surface Temperature Dependence}} \label{fail}

\begin{figure}[h]
\begin{center}
\includegraphics[width=0.50\textwidth]{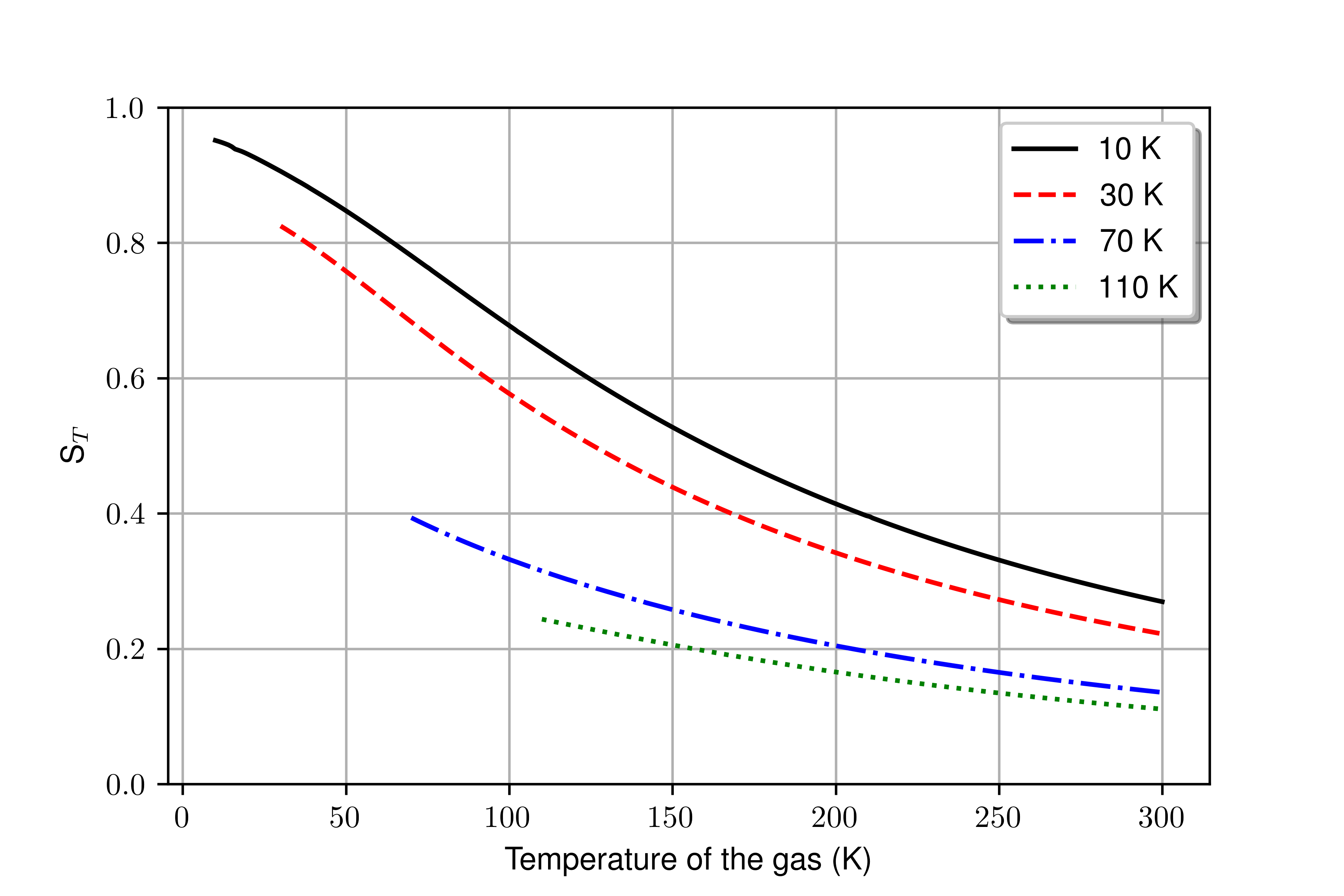}
\caption{Sticking coefficients ($S_\text{T}$) at different surface temperatures. Values are shown for $T_\text{gas} \ge T_\text{sur}$.}
\label{figfail}
\end{center}
\end{figure}

We also studied the dependence of the sticking probability on the surface temperature. For doing so we employed different surface temperatures, namely 30~K, 70~K and 110~K additionally to the 10~K reported above. Total simulation times were increased to 15~ps. As a consequence of the increased simulation times we sampled fewer incident energies (1, 3, 10, 15, 20, 25, 30, 50 \& 60 meV). For the angular sampling we used 14 $\theta$-values and 18 $\phi$-values for a surface temperature of 30 K, and 14 $\theta$-values and 14 $\phi$-values for 70~K and 110~K. The purpose of the increased sampling time is to investigate if that improves the description of surface-to-adsorbate energy transfer. Results for $S_\text{T}$ can be found in \figref{figfail}.

It is important to remark that for temperatures \textgreater 10 K, the values provided for $S_\text{T}$ constitute only an upper bound approximation to the experimental values. Surface-to-adsorbate energy transfer can lead to thermal desorption long after the collision event.  The bouncing or sticking of a molecule on the surface is greatly dependent on the redistribution of energy between the degrees of freedom of molecule and those of the surface after collision. Within these processes, surface-to-adsorbate energy transfer gets more important at higher surface temperatures, because more thermal energy is available. Thermal energy redistribution leading to thermal desorption is a slow process.
If we assume that desorption has to overcome an energy barrier of 300~K (the average binding energy shown  in \figref{binding_fig}), the half-life of H$_2$ on the surface would be tens of nanoseconds at 30~K. This is clearly longer than the duration of an adsorption event followed by thermalization (picoseconds). This estimate is based on an assumed Arrhenius behavior of the thermal desorption rate constant,  $k=\nu\exp(-E_\text{bin}/T)$, with $\nu= 100$ cm$^{-1}$, as a frequency estimate of a vibration in a weakly bound system. Thus, sticking and thermal desorption should be considered separately because of their separation in time scale. At 10 K though, this is not a relevant factor, because the available thermal energy is very low, and the sticking probability is dominated mostly by the transfer of energy from the adsorbate to the surface, a process that is heavy localized in the collision point.

The fact that no sticking of H$_{2}$ on ASW is observed above a certain thermal desorption temperature should be interpreted independently of the sticking process. H$_{2}$ may stick on the surface at higher temperature, as found in our simulations, but will desorb before being detected. In the case of H$_{2}$/ASW the thermal desorption temperature is particularly low, as shown in the experiments accounting for surface temperature by He et al. \cite{He2016}. 

\section{Discussion}

\subsection{Comparison with other works}

In this section, we compare our results to those from the literature. The focus lies on two experimental measurements of sticking of H$_2$ on ASW by Matar et al.\ \cite{Matar2010}, in the following referred to as MATAR2010 and by He et al.\ \cite{He2016}, HE2016. 

The two studies used different methods to quantify the sticking. MATAR2010 measured the sticking coefficient of \ce{H2} on non-porous amorphous solid water as a function of the beam temperature, keeping the surface temperature  constant at 10 K. They employed the King--Wells method to calculate sticking coefficients. By contrast, HE2016 studied the influence of the surface temperature on the same system, with a beam estimated to be at 300 K, obtained from the fit of the time-of-flight spectra of \ce{O2} molecules. In addition to that, they argued that the King--Wells method is not necessarily suitable for measuring sticking coefficients at cryogenic temperatures, because of the lack of time resolution and excess of pumping due to condensation in some parts of the apparatus. We refer the reader to their interesting discussion \cite{He2016}. 

Whatever the reason, both studies can be directly compared for only one temperature, 10 K for the surface and 300 K for the beam, given the different factors under consideration. In that particular case their results differ significantly, MATAR2010 find a sticking coefficent of 0.22, while HE2016 obtained 0.58. This discrepancy is argued in HE2016 to be a result of the employment of the King--Wells technique with its associated drawbacks. 

\begin{figure}[h]
\begin{center}
\includegraphics[width=0.50\textwidth]{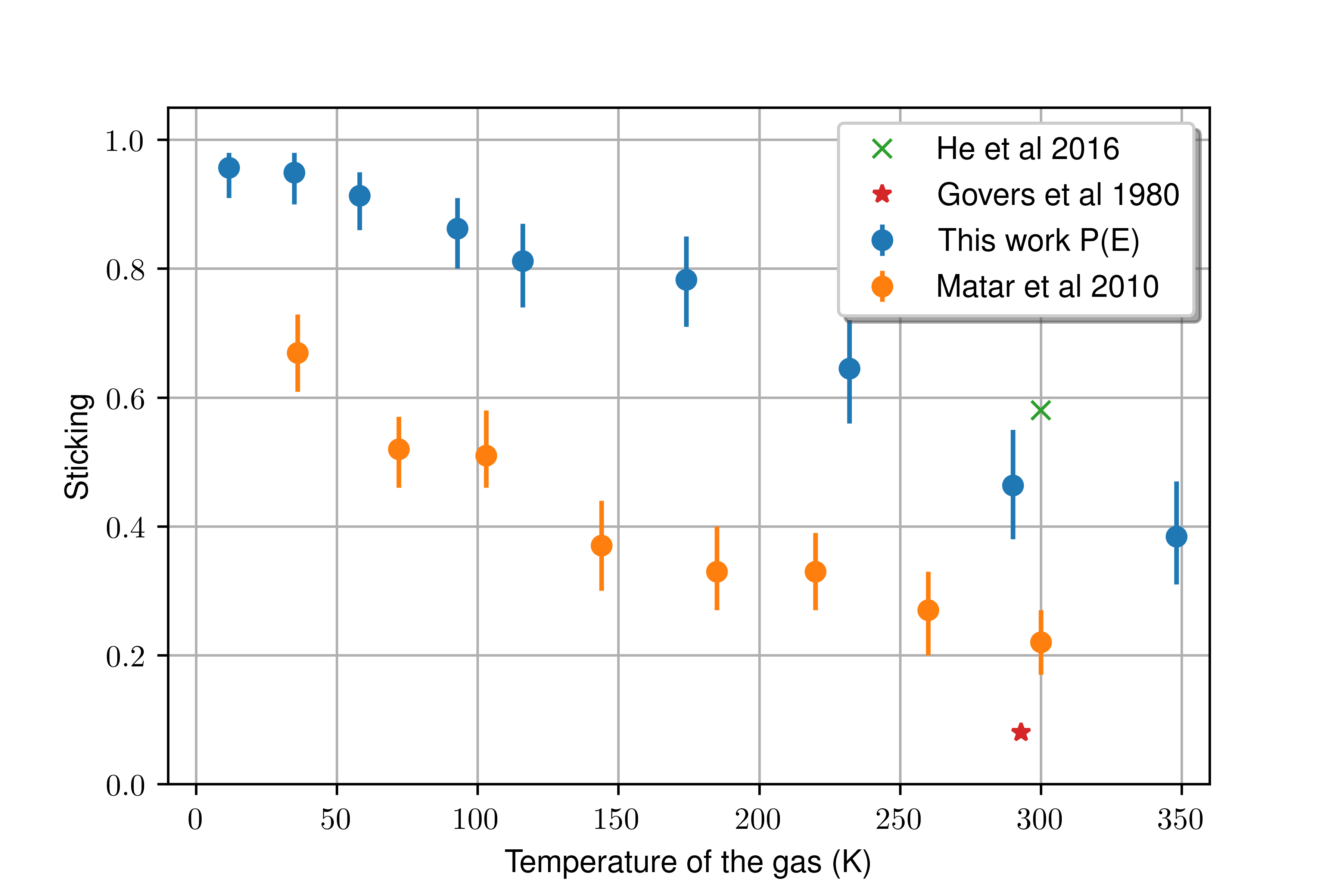}
\caption{Comparison between the sticking probabilities $P(E)$ of this work and the results by Matar et al.\ 2010 \cite{Matar2010}, He et al.\ \cite{He2016}, and Govers et al.\ \cite{Govers1980} at a surface temperature of 10 K.}
\label{lastimage}
\end{center}
\end{figure}

\figref{lastimage} compares our results to the literature data. We compare the literature values to $P(E)$ rather than $S_\text{T}$ because of the experimental physical conditions of an effusive molecular beam, in which the beam is not in thermal equilibrium and the beam temperature is a representation of the energy of the particles \cite{Matar2010}. We have also included a comparison with the classical work by Govers \cite{Govers1980}. A qualitative comparison shows that our values are in good agreement with the more recent values by HE2016 within our error bars. We have not found an error estimate in HE2016, but we assume an error comparable to ours. We can, however, compare our results at only one particular temperature, since the work by HE2016 focuses on the study of the influence of the surface temperature, fixing the beam temperature.  The comparison with MATAR2010 is more straightforward since they provided the energy dependence that is directly comparable with our simulations, showing in all cases a lower sticking coefficient, even considering error bars. This is in accordance with the arguments against King--Wells method for measuring sticking coefficients.


Finally, concerning our surface-temperature dependence simulations, we are not in a position to compare them with the results provided in HE2016, for the reasons stated in section \ref{fail}: sticking and thermal desorption are two different processes with rather different time scales. We can only simulate the former, while both influence the experiment.

\subsection{Strengths and Weaknesses of our method}

In order to gauge the predictive power of our method for the determination of adsorption dynamics, we discuss its advantages and disadvantages as currently apparent. The approach is generally applicable to any kind of surface, amorphous or crystalline. Our quantum-chemistry based underlying potential allows to study chemisorption, chemical reactions upon adsorption with surface atoms as well as pre-adsorbed species (Eley--Rideal mechanism). This is not relevant for H$_2$-adsorption, but may be important for other species. We have seen that a certain amount of sampling, in the order of $10^6$ energy and gradient evaluations, is necessary. This can be achieved by a QM/MM separation with a fast QM method. Our combination works well for H$_2$ on ASW, but other electronic structure methods may be necessary for other adsorbate/surface combinations. Interatomic potentials for our adsorbate-surface pair are calculated \textit{on-the-fly}, and do not require a system-specific force-field parameterization or potential energy surface fit. The range of energies that can be considered for the adsorbate is broad enough to allow the study different processes with relevance to astrophysics, ranging from physisorption to processing with high energy particles.

Certainly, there are limitations, which could be kept under control for the H$_2$/ASW system but may become more severe in other combinations. The limitations in CPU time require a limitation in the structural model, i.e.\ a surface model of finite size. The QM/MM approach requires all direct interactions, especially the first impact, to happen within the QM region. Especially for a rough surface, that means that the angle of incidence is limited. We restricted it to $\theta\le 65$\degree. However, we found that the dependence of the sticking probability on the angle is limited, so that restriction may be minor. However, surface roughness in the microscale or larger, especially pores, may lead to multiple bounces, in which the adsorbate loses energy with each bounce. That will increase the sticking probability compared to our values, in which we always observe single or double bounces. 

In contrast to force-field-based sampling approaches, our quantum-chemistry-based approach does not allow to partition the potential energy between individual atoms. Thus, we can not use the potential energy of the adsorbate as criterion for a categorization of an event as sticking event. Such a categorization would allow a decrease in simulation time and could, in principle, help in the determination of the influence of the surface temperature.

\begin{figure}[h]
\begin{center}
\includegraphics[width=0.50\textwidth]{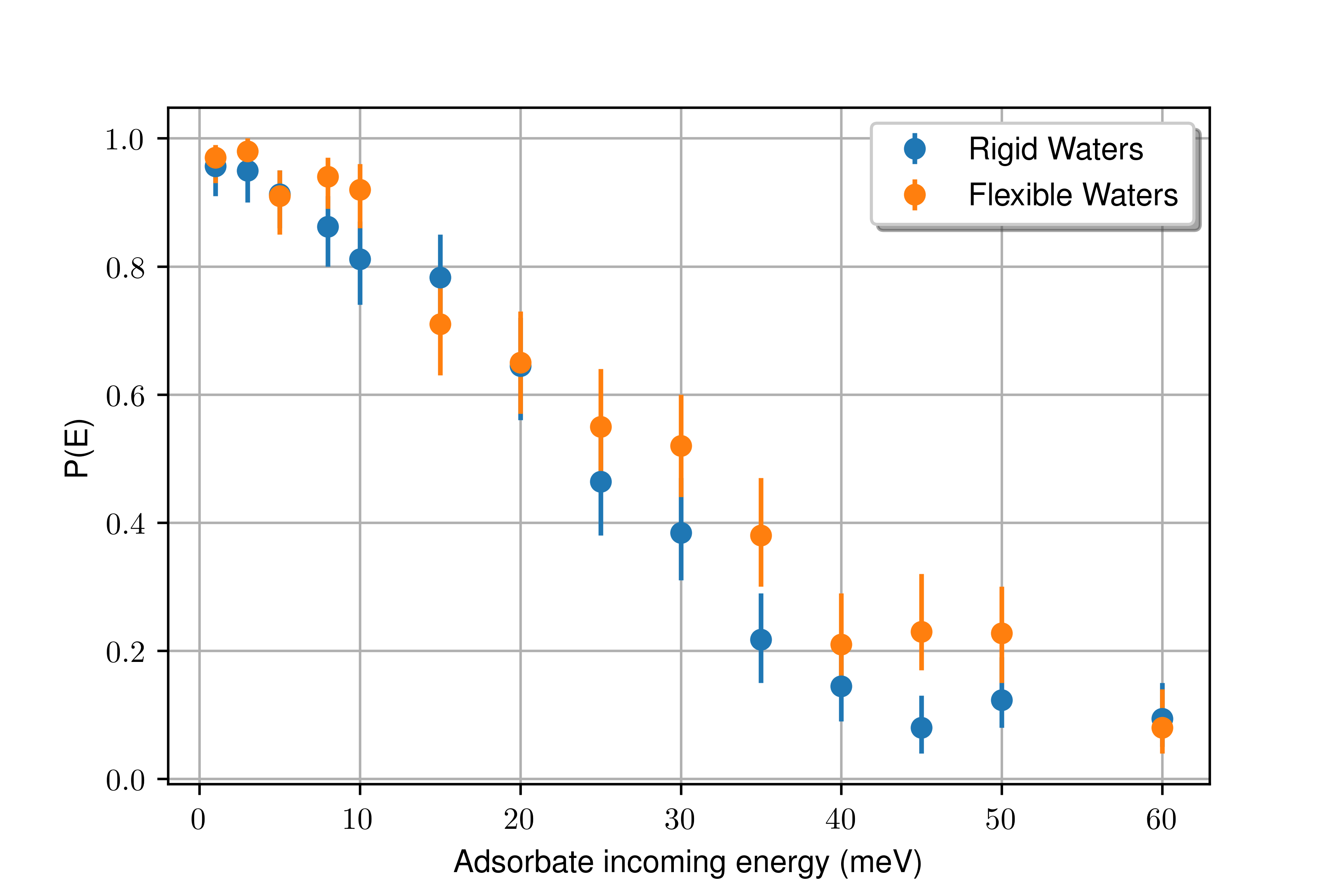}
\caption{Comparison between the sticking probabilities $P(E)$ at a surface temperature of 10 K employing a rigid water surface model for the quantum part versus a model employing flexible quantum waters.}
\label{rvf}
\end{center}
\end{figure}

While electrons are being treated quantum-mechanically in our simulations, the dynamics of the nuclei is classical. In particular this means that rotations and vibrations, which are quantized in reality, can be excited by arbitrarily small energy values in our simulations. Especially the excitation of internal vibrations of the water molecules (and H$_2$), which require comparably large energy quanta in reality, may have an influence on the heat dissipation following a sticking event. The impact of this approximation is difficult to assess without quantum dynamics, which is infeasable for the given system size. One simple test is to compare a flexible water model to an internally rigid one. The rigid water model is chosen to represent a molecule in which internal motion can not be excited at all. We calculated sticking probabilities for both and found them to be very close in their behavior within our confidence interval, especially at low incoming energies. See \figref{rvf} for a comparison at a surface temerature of 10 K.  Thus, we claim that the classical nuclear dynamics is a good approximation.

We find finite ($>0$) sticking probabilities at surface temperatures above the experimental thermal desorption temperature. We interpret this as a physically meaningful process: even at elevated temperature, there is a finite probability for molecules to be adsorbed, even though they will be thermally desorbed before a significant surface coverage can be reached. The time scales for adsorption and thermal desorption are rather different. This may have important implications for astrochemistry: even at temperatures above the thermal desorption energy there is a finite probability for surface reactions, since a thermally activated adsorbate is bound to the surface. This possibility should be tested experimentally with fast-reacting adsorbates, for which a well-defined desorption temperature is still measurable. 

\section{Conclusions}

We provided a new set of sticking coefficients for the \ce{H2}/ASW system that we consider of special relevance for astrochemical modeling at low gas temperatures such as the ones found in an interstellar dense cloud. The method we have employed is based on quantum-chemistry-based molecular dynamics simulations, instead of force-field-based dynamics used previously. The main points we learn from the study are:

\begin{enumerate}

\item We determined the binding energy distribution of \ce{H2} in ASW for our model. The distribution is quite broad, but centered around a binding energy of 300~K. The results agree with the experimental data available and are coherent with the theoretical results already published for the \ce{H}/ASW system.

\item Our sampling strategy and simulation conditions constitute a general way to study collision dynamics of an adsorbate with a surface and relies on random lateral displacements and selection of different starting configurations to sample amorphous surfaces.

\item GFN2-xTB as electronic solver provides similar accuracy as high-level ab-initio calculations for the \ce{H2}/\ce{H2O} system, specially at long distances. 

\item Sticking coefficients depend strongly on the adsorbate kinetic energy. This dependence is well-resolved in our study. We provided values for the sticking coefficients using a range of initial adsorbate energies. The sampling includes nearly 2000 molecular dynamics runs in order to obtain sufficient statistics. The sticking coefficient always lower than unity in our simulations.

\item The approximation of the sticking probability as depending exponentially on the energy, quite common in the literature for hydrogen atom adsorption studies, does not entirely hold for the \ce{H_{2}} molecule, providing too low values of the sticking coefficient at low and medium gas temperatures. 

\item We considered the dependence of the sticking coefficient on the angle of incidence. The overall influence of the angle is lower than that of the incoming energy and surface temperature.

\item We observed sticking above the experimental thermal desorption temperature. The time scale for sticking and thermalization vs. thermal desorption are rather different. We conclude that these processes should be treated separately and that adsorbates may be available on the surface even above the thermal desorption temperature.

\item Comparison with experiments have been made showing a good degree of accuracy with the latest published results. However, a direct comparison is rarely possible, as sticking coefficients depend on the conditions of experiments and simulations. We emphasize that the evaluation of these magnitudes would greatly benefit from cooperative work between theory and experiment.

\end{enumerate}

The method we have developed is general for every possible surface-adsorbate pair provided an accurate interaction potential. We are currently working on the study of the adsorption of molecules for which no sticking coefficient is reported in the literature so far, including charged species and radicals, with description of chemisorption and in-situ surface reactions. Further modifications for the method are available, including a better description of the potential of the molecular mechanics part of the surface and the construction of other astrophysically relevant surfaces.

\section*{Acknowledgements}
We would like to acknowledge funding from the European Union's Horizon 2020 research and innovation programme (grant agreement No. 646717, TUNNELCHEM). We also like to also acknowledge the support with computer time by the state of Baden-W\"{u}rttemberg through bwHPC and the German Research Foundation (DFG) through grant no. INST 40/467-1FUGG. G.M would also like to thank the Alexander von Humboldt foundation for a recently obtained postdoctoral fellowship.

\bibliography{ref} 
\bibliographystyle{unsrt}

\end{document}